\documentclass[floatfix,lengthcheck,showpacs,amssymb,amsmath,amsfonts,twocolumn,nofootinbib,longbibliography,superscriptaddress]{revtex4-1}
\usepackage{epsfig}
\usepackage{bm}
\usepackage{latexsym}
\usepackage{natbib}
\usepackage{url}
\usepackage{orcidlink}
\usepackage[T1]{fontenc}
\usepackage{lmodern}
\usepackage{amsfonts,amssymb,amsmath}
\usepackage{graphicx} 
\providecommand{\bibinfo}[2]{#2}

\usepackage{psfrag}
\usepackage{enumitem}
\usepackage{float}
\usepackage{mathrsfs}
\usepackage[dvipsnames]{xcolor}
\usepackage[utf8]{inputenc}
\usepackage{comment}
\usepackage{wrapfig}
\usepackage{balance}
\usepackage{amssymb}
\hypersetup{
  colorlinks=true,
  linkcolor=red,   
  citecolor=red,    
  urlcolor=red,        
  filecolor=red         
}
\usepackage{tabularx}
\usepackage{subcaption}
\usepackage{booktabs}
\usepackage{array}
\usepackage{cleveref}
\allowdisplaybreaks
\begin{document}

\title{Constraints on Line-of-Sight Acceleration from O1-O4}

\author{Labani Roy\,\orcidlink{0009-0004-8445-6212}}
\email{lroy02@syr.edu}
\affiliation{Department of Physics, Syracuse University, Syracuse, NY 13244, USA}

\author{Alexander H. Nitz\,\orcidlink{0000-0002-1850-4587}}
\affiliation{Department of Physics, Syracuse University, Syracuse, NY 13244, USA}
\date{\today}

\begin{abstract}

A compact binary will experience a center-of-mass (CoM) acceleration in the vicinity of a massive third object. The line-of-sight (LOS) component of this acceleration is imprinted on gravitational waves (GWs) produced by the compact binary as a time-varying Doppler shift. The observation of a non-zero LOS acceleration may indicate the binary is in a dense environment, such as an active galactic nucleus (AGN) disk or nuclear star cluster, etc. We measure the LOS acceleration of all compact binaries observed through the first part of the fourth observing run (O1-O4a) of Advanced LIGO and Virgo in addition to select binaries from later observing runs. We introduce a new method to model the LOS acceleration by directly applying the time-varying Doppler shift in the time domain to the signal produced in the binary's frame; this method can be applied to any waveform model including those with higher order modes, eccentricity, and precession. We find the LOS acceleration for all known binaries to date is consistent with zero. We find that the effects of eccentricity and LOS acceleration are partially degenerate as observed in binaries such as GW200105. Current ground-based observatories are sensitive enough to only constrain scenarios that produce high accelerations, e.g $\sim 10^{-2~}(10^{-6})~\textrm{c}/s$ for binary black hole (BBH) (binary neutron star (BNS)) sources; however, next-generation observatories may be able to constrain the accelerations expected in some dense environments. 

\end{abstract}

\maketitle

\section{Introduction}
Since 2015, from the first observing run (O1) to the fourth observing run (O4), the Laser Interferometer Gravitational-Wave Observatory (LIGO) \cite{LIGOScientific:2016aoc, LIGOScientific:2014pky} and Virgo \cite{VIRGO:2014yos} detector network have observed $\mathcal{O}(100)$ GW signals from compact binary mergers \cite{Nitz:2021zwj, KAGRA:2021vkt, LIGOScientific:2025slb, LIGOScientific:2026sit}. GW170817 \cite{LIGOScientific:2017vwq} and GW190425 \cite{LIGOScientific:2020aai} are  BNS mergers detected during the second and third observing runs, respectively. GW170817 is the first GW event associated with an electromagnetic counterpart \cite{LIGOScientific:2017ync, LIGOScientific:2017zic}. So far, four neutron star–black hole (NSBH) merger events have been observed: GW200105, GW200115 \cite{LIGOScientific:2021qlt}, GW230518, and GW230529 \cite{LIGOScientific:2025slb, LIGOScientific:2025pvj}. Apart from these BNS and NSBH events, the detector network has also detected numerous BBH mergers. Beyond estimating the source properties of these compact binaries, probing and constraining their host environments has become an active area of research \cite{Mapelli:2020vfa, CanevaSantoro:2023aol}.   
The environments surrounding compact objects influence their formation, dynamics, and evolution. Compact binaries formed in dense stellar environments, such as globular clusters \cite{Ziosi:2014sra, Rodriguez:2015oxa, Mapelli:2021gyv, Mapelli:2021taw}, nuclear star clusters \cite{Mapelli:2021taw, 2022MNRAS.517.2953T, DiCarlo:2020lfa}, and active galactic nucleus (AGN) disks \cite{Sedda:2023big, 2023MNRAS.524.2770R, Ishibashi:2020zzy, Mapelli:2021taw}, can undergo frequent dynamical interactions that significantly affect their orbital properties \cite{CanevaSantoro:2023aol, Li:2025wia}. Such environmental effects may also lead to the electromagnetic counterparts associated with GW signals from BBH mergers occurring in non-vacuum environments \cite{Loeb:2016fzn, McKernan:2019hqs, Perna:2019pzr, Graham:2020gwr, Li:2025wia}.

Considerable effort has been devoted to understanding the immediate environments and formation pathways of compact binaries. \citet{Rodriguez:2016vmx} explored the spin-tilt distribution of BBHs formed in isolated and globular clusters and found that higher values for the negative spin-alignment indicate that the BH is formed in dense environments. In isolated binary evolution, spins are preferentially aligned to the orbital angular momentum, resulting in a positive effective spin parameter \cite{Kalogera:1999tq}. Moreover, hierarchical mergers in clusters or AGN disks can produce remnants with high spin magnitudes, providing an additional diagnostic of repeated mergers \cite{Tagawa:2021ofj, ArcaSedda:2018cyl}. Low-metallicity progenitors produce more massive black holes, as shown by population synthesis studies \cite{Belczynski:2016obo, Mapelli:2017hqk}. 
\citet{Zevin:2017evb} showed that solely with the chirp mass measurements, it is possible to constrain the natal kick prescriptions and the relative fraction of systems originating from different formation channels. 
Eccentricity is a potential probe of a binary's formation channels \cite{Mapelli:2021taw, Romero-Shaw:2019itr, Zevin:2021rtf, Fumagalli:2024gko, Saini:2023wdk, Morras:2025xfu}. Binaries assembled dynamically in globular clusters, galactic nuclei, or hierarchical triples can retain measurable eccentricity at frequencies accessible to LIGO or future detectors with improved sensitivity \cite{Lower:2018seu, Saini:2023wdk, Randall:2018qna, Samsing:2018isx}. Recent work \cite{Giri:2026wgy} has shown that brief three-body encounters can produce a distinctive morphology characterized by dephasing and amplitude modulation in the GW waveform, providing a probe of compact binaries embedded in dense environments. 

\begin{figure}[]
    \begin{center}
\includegraphics[width=10cm,height=9cm,angle=0]{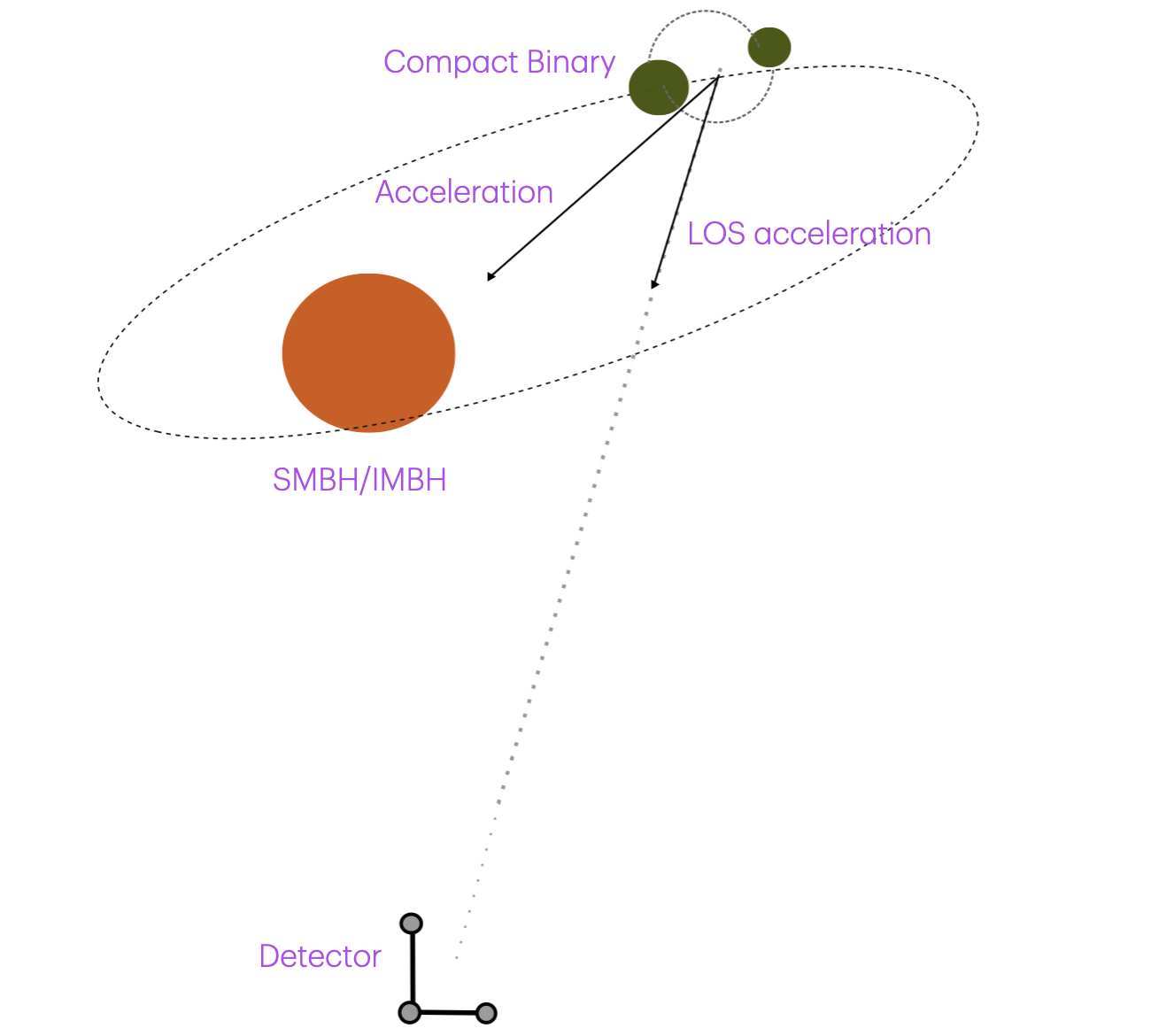}
    \caption{Schematic diagram of a possible environment where a binary is in the vicinity of an SMBH or IMBH. The presence of an SMBH/IMBH creates a net acceleration on the binary's CoM. The LOS component of this acceleration will modulate the GWs emitted from the binary, which can be detected by GW observatories.}
    \label{fig-system}
    \end{center}
\end{figure}

A number of methods have been used to understand how binaries form and the environments in which they reside. In this work, we investigate another approach to assessing the environments of compact binaries. Identifying the host environment of compact binaries is challenging because most GW sources, particularly BBHs, lack electromagnetic counterparts, and their sky localisation is often poor in current detector networks \cite{Tiwari:2024pvb}. Nevertheless, some astrophysical environments can leave measurable imprints on the observed GW signal. For example, binaries embedded in an AGN disk or orbiting near a supermassive black hole (SMBH) or intermediate-mass black hole (IMBH) are expected to experience significant CoM accelerations due to the gravitational influence of the third body, whereas binaries formed through isolated binary evolution are expected to have much weaker CoM accelerations \cite{Inayoshi:2017hgw, Vijaykumar:2023tjg, Tamanini:2019usx}. Previous studies have shown that such CoM motion can leave measurable imprints on the GW signal, with eccentric outer orbits around a third body introducing characteristic features in the phase evolution \cite{Hendriks:2024zbu}, while dynamical perturbations during three-body encounters and R\o mer delay effects can further modify the observed waveform \cite{Samsing:2024syt}. \citet{Tiwari:2024pvb} argued that constraints on the CoM motion of compact binary coalescences (CBC) can serve as a complementary probe of their formation environments, since different environments are expected to induce different CoM kinematics. Unlike a constant CoM velocity (e.g., due to cosmological expansion), which produces a uniform redshift that is degenerate with the binary's total source frame mass, a CoM acceleration induces a time-varying Doppler shift. The LOS component of the CoM acceleration contributes to this time-varying Doppler shift, producing a time-dependent frequency evolution that accumulates as an additional phase shift in the observed GW signal \cite{Tamanini:2019usx}. The magnitude of this phase drift depends on the mass of the third object and its separation from the compact binary \cite{Vijaykumar:2023tjg}. Fig~\ref{fig-system} illustrates one possible scenario in which a compact binary orbits in the vicinity of a central SMBH or IMBH.

Previous waveform models that account for LOS acceleration have largely been developed within the post-newtonian (PN) \cite{Blanchet:2013haa} framework. \citet{Yunes:2010sm} and \citet{Bonvin:2016qxr} showed that the leading order phase correction in phase appears at $-4$ PN order due to LOS acceleration. \citet{Tamanini:2019usx, Vijaykumar:2023tjg} further added phase correction to the leading order term. \citet{Yang_2025} used 3.5 PN phase correction and claimed the presence of non-zero acceleration in GW$190814$ \cite{LIGOScientific:2020zkf}. Recent studies extended the PN phase correction framework for systems including higher-order modes and eccentricity \cite{Roy:2026mco}.  

In this work, we adopt the same physical approximation (constant LOS acceleration) as used in previous studies. Instead of using the PN \cite{Blanchet:2013haa} approximation, we applied a time-varying Doppler transformation directly on existing waveform model in time-domain. As a result, our approach is compatible with any underlying waveform model, including those incorporating higher-order modes, eccentricity, and spin precession. The method is applicable whenever the assumption that the entire binary experiences a uniform acceleration over the duration of the observed signal is valid. We estimate the LOS acceleration for GW sources through O4a \cite{LIGOScientific:2025slb} and a subset from later observing runs \cite{LIGOScientific:2026sit}. We observe a correlation between eccentricity and LOS acceleration in events such as GW200105 \cite{LIGOScientific:2021qlt}. For GW190814 \cite{LIGOScientific:2020zkf}, we find no evidence for non-zero LOS acceleration, consistent with recent studies \cite{Hendriks:2026kys, Pathak:2026cik, Roy:2026mco}. Our analysis shows that all the observed events are consistent with zero LOS acceleration. Furthermore, the LOS accelerations expected for most astrophysically motivated scenarios lie well below the sensitivity of current ground-based detectors. However, we expect the next generation ground based observatories and space-based observatories may able to constrain LOS accelerations for some likely astrophysical cases.

\begin{figure*}
        \centering
        \includegraphics[width=\linewidth]{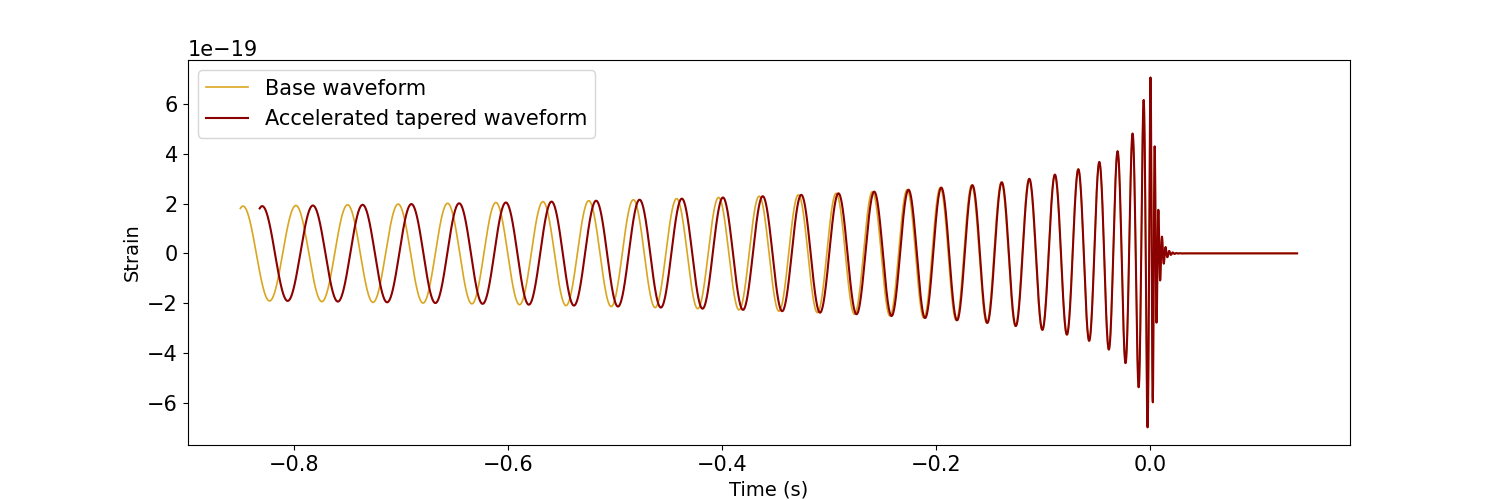}
        \caption{Comparison of the GW strain from an \texttt{EOBNRv2HM} \cite{Pan:2011gk} and its Doppler shifted counterpart for a $30 M_{\odot}$ and $30 M_{\odot}$ binary starting at 20 Hz. A LOS acceleration of $0.05 \,\textrm{c}s^{-1}$ is applied to the waveform, which compresses the inspiral and stretches the ringdown portion.} 
        \label{fig-step2}
    \label{fig-waveformmodel}
\end{figure*}

\section{Modeling Accelerating Binaries}\label{II}

The binary's CoM acceleration induced by its environment modifies the frequency evolution of the GW signal. The observed GW signal experiences a time-varying Doppler shift due to the LOS component of CoM acceleration.  
Previous studies used waveform models based on the PN \cite{Blanchet:2013haa} expansion to introduce LOS acceleration as a time-varying Doppler shift by adding additional phase correction to this expansion. A common assumption is that the binary undergoes a uniform acceleration. 
The resulting GW phase correction due to LOS acceleration enters at $-4$PN order \cite{Yunes:2010sm, Bonvin:2016qxr}. This leading-order correction has been employed in recent analyses of GW events \cite{Hendriks:2026kys, Pathak:2026cik}. Higher-order corrections have subsequently been derived up to 3.5 PN order \cite{Tamanini:2019usx,Vijaykumar:2023tjg} and incorporated into frequency domain waveform models such as \texttt{IMRPhenomXPHM} \cite{Vijaykumar:2023tjg, Yang_2025,Garcia-Quiros:2020qpx}. More recently, \citet{Roy:2026mco} implemented LOS acceleration effects consistently in frequency-domain waveform models, incorporating the PN phasing contributions up to 3.5PN order, mode-dependent amplitude corrections, higher harmonics, and precession and eccentricity. Concurrent with this work, an independent study by \citet{Pompili:2026tdf} developed a related time-domain implementation using \texttt{SEOBNR} \cite{Ramos-Buades:2023ehm, Gamboa:2026jht} waveform models and jointly analyzed LOS acceleration and orbital eccentricity.

In this work, we apply the time-varying Doppler transformation directly to the GW signal in the CoM frame via time shifting and interpolation. Because the signal model is directly transformed from the CoM frame, which is being accelerated to an observer in an inertial reference frame as observed on the Earth, all amplitude and phase corrections are implicitly accounted for. The model nevertheless retains the standard assumptions of earlier PN \cite{Blanchet:2013haa} expansions: the binary undergoes uniform acceleration during the observation, while tidal perturbations from the tertiary object and other relativistic effects (e.g., lensing \cite{Bartelmann:2010fz}) are neglected. 
The uniform acceleration approximation is valid only when the observation duration is sufficiently short that the acceleration does not vary appreciably over the signal. This assumption can also break down for compact binaries orbiting very close to a third body, where tidal effects and higher-order acceleration effects become important. 

We use the following steps to generate the time-varying Doppler-shifted waveform :
\begin{itemize}
    \item A pre-existing (base) waveform model is used to generate signal in the CoM frame. 
    \item The time-varying Doppler transformation is applied to the base waveform model by shifting its time samples (see Eq.~\ref{model}). A non-relativistic truncation ($v/c<0.1$) is also applied on top of this signal.
    \item An equally sampled time series is generated by using third-order spline interpolation. At the very end, 80 ms of constant tapering is applied at the start of the time series.
\end{itemize}

\begin{align}
    f_{\mathrm{acc}} & = \frac{1}{1 + \frac{v}{c}} f_{\mathrm{non-acc}} \nonumber\,\\
    \implies \frac{d\phi}{dt}\Bigg|_{\mathrm{acc}} & = \frac{1}{1 + \frac{v}{c}} \frac{d\phi}{dt}\Bigg|_{\mathrm{non-acc}} \nonumber\,\\
    \implies dt_{\mathrm{acc}} & = \left(1 + \frac{v}{c}\right) dt_{\mathrm{non-acc}} \nonumber\, \\
    \implies dt_{\mathrm{acc}} & = \left(1 + \frac{v_0}{c} + \frac{at}{c}\right) dt_{\mathrm{non-acc}}\,
    \label{model}
\end{align}

 Eq.~\ref{model} shows the relation between the Doppler shifted time interval $(dt_{\mathrm{acc}})$
 to the time interval when there is no acceleration $(dt_{\mathrm{non-acc}})$. In Eq.~\ref{model}, $v =$ LOS velocity, $a =$ LOS acceleration, $c =$ speed of the light. As prior studies, we use a simple $v = v_0 + \frac{dv}{dt}t$ relation, where $dv/dt = a$. 
 In general, any profile for $v$ (example, $v = v_0 + \frac{dv}{dt}t + \frac{d^2v}{dt^2}t+...)$ can be used in this waveform model. 
Since an overall constant CoM velocity is degenerate with the cosmological redshift and the total binary mass, the value of $v_0$ can be absorbed into these parameters. In this work, without loss of generality, we fix our definitions so that $v_0$ is defined as zero at the coalescence time. With this convention, a positive (negative) acceleration implies that the binary is accelerating away from (toward) the observer. So a positive (negative) acceleration will stretch (squeeze) the ringdown portion of the signal and compresses (stretches) the inspiral part relative to the base model. Fig.~\ref{fig-step2} shows an example comparing a waveform with acceleration effects applied to the underlying base waveform. 

This method requires conversion between the frequency and time domains, introducing an $\mathcal{O}(N \log N)$ cost. The subsequent Doppler remapping and spline interpolation steps scale approximately linearly with the number of samples. These steps therefore contribute a negligible computational cost relative to base-waveform generation.

A closely related approach was developed by \citet{Chamberlain:2018snj}, where a frequency-domain waveform is transformed to the time domain, the time-varying Doppler remapping is applied through an expansion in the small velocity parameter, and the waveform is transformed back to the frequency domain.

\subsection{Accuracy of the waveform model}

\begin{figure*}
    \centering
    \includegraphics[width=0.8\textwidth,height=6cm,keepaspectratio]{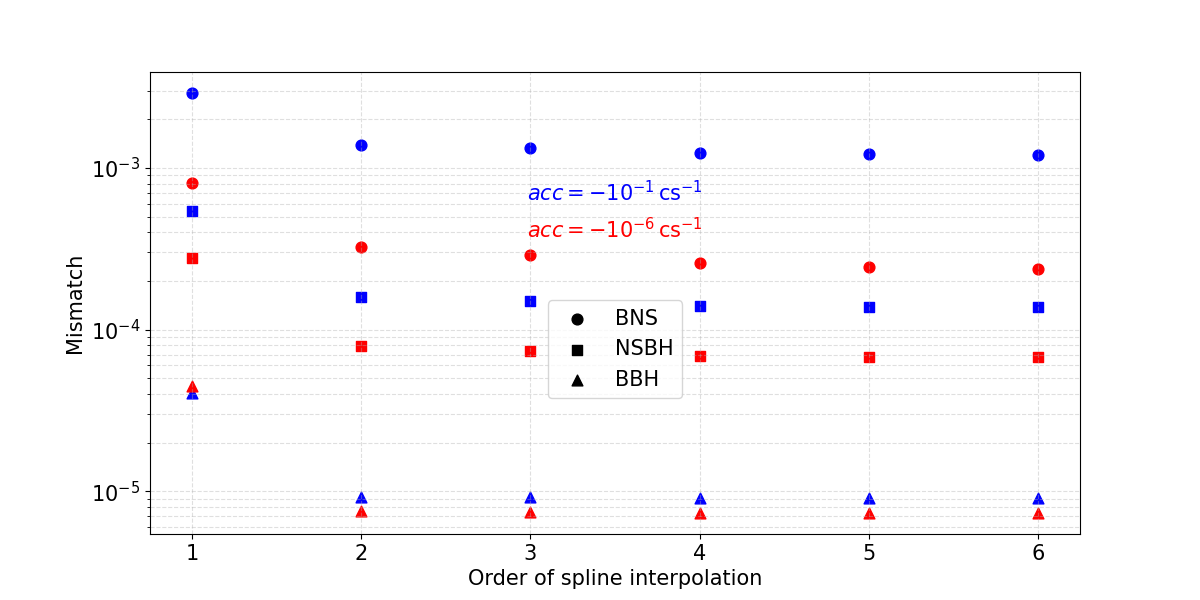}
    \caption{This figure shows the mismatch between our continuum-limit waveform and a waveform generated with a sampling interval of 1/4096 as a function of interpolation order. Red and blue denote two reference accelerations $-10^{-6}\,\textrm{c}s^{-1}$ and $-10^{-1}\,\textrm{c}s^{-1}$, respectively. Results are shown for a representative BNS (circles), NSBH (squares), and BBH (triangles) system. Mismatch decreases with the interpolation order. Most of these accuracies are constrained by the sample-rate not by the interpolation order. The interpolation order at third/fourth order is sufficient for mismatches down to $10^{-6} (10^{-4})$ for $-10^{-6}\,\textrm{c}s^{-1}$ at a sample-rate of $1/4096$ for BBH (BNS).}
    \label{fig-interpolation}
\end{figure*}

\begin{figure*}[]
    \centering
    \includegraphics[width=0.8\textwidth,height=6cm,keepaspectratio]{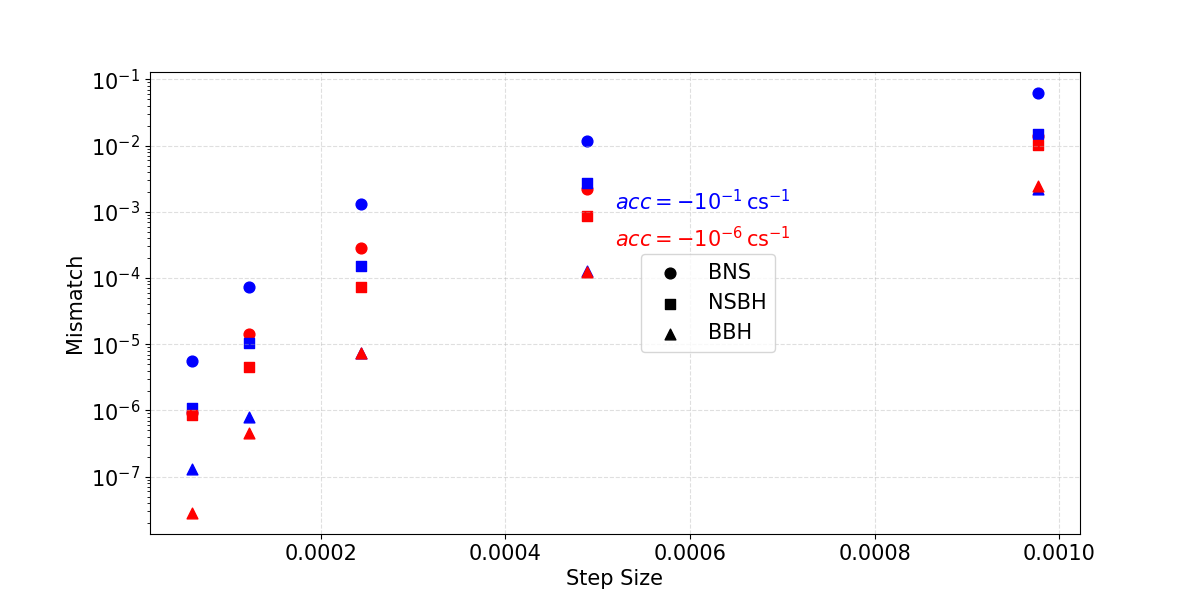}
    \caption{This figure shows the mismatch between our continuum-limit waveform and a waveform generated  with third order spline interpolation as a function of the step size. Red and blue denote two reference LOS accelerations $-10^{-6}\,\textrm{c}s^{-1}$ and $-10^{-1}\,\textrm{c}s^{-1}$, respectively. Results are shown for representative BNS (circles), NSBH (squares), and BBH (triangles) systems. As the step size decreases, interpolation error becomes negligible. At a sample rate of $1/4096$, our mismatch is not dominated by current waveform systematics. For the range of accelerations that is applicable to each of these sources, the third order spline interpolation is sufficiently accurate, so that the expected PE will be dominated by the waveform and statistical systematics, not by the interpolation order.}
    \label{fig-sample-rate}
\end{figure*}

\begin{figure*}[]
    \centering
    \includegraphics[width=0.8\textwidth,height=6cm,keepaspectratio]{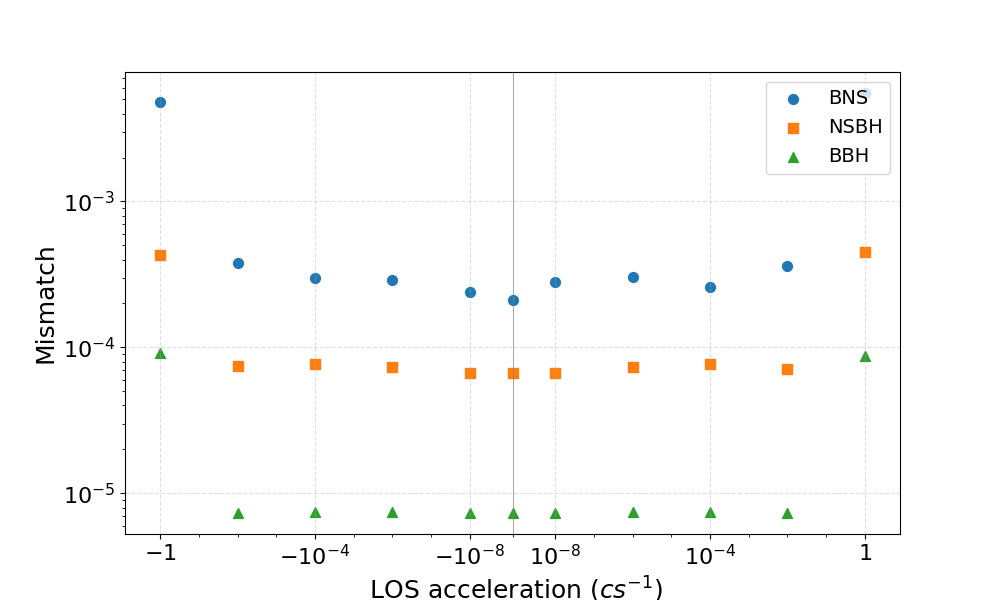}
    \caption{This figure shows the mismatch between our continuum-limit waveform and a waveform generated with a sampling interval of $1/4096$ as a function of the LOS acceleration. Circles, squares, and triangles denote representative BNS, NSBH, and BBH systems, respectively. The mismatch increases with increasing LOS acceleration for all binary types.}
    \label{fig-acceleration}
\end{figure*}

The final step of our procedure is to interpolate the time-shifted samples to produce a new waveform at a fixed sample rate. This is required for compatibility with standard analysis techniques \cite{Cooley:1965zz} and parameter estimation (PE) tools \cite{Usman:2015kfa, Ashton:2018jfp, lalsuite}.
We verify that the interpolated signals accurately represent the transformed waveform by comparing them to a high-resolution reference signal that approximates the continuum limit.

To quantify the difference between the continuum waveform ($h_1$) and a coarsely sampled ($h_2$) one, we calculate the mismatch, which is defined as,

\begin{align*}
    1 - \left\langle h_1, h_2 \right\rangle & = 1 - \max_{\phi_c, t_c}\frac{\left( h_1, h_2 \right)}
{\sqrt{\left( h_1, h_1 \right)\left( h_2, h_2 \right)}}
\end{align*}

The inner product $\left( h_1, h_2 \right)$ is defined as,
 \begin{equation*}
     \left( h_1, h_2 \right) = 4\Re \int_{f_{\min}}^{f_{\max}}
\frac{\tilde{h}_1(f) \tilde{h}_2^{*}(f)}
{S_n(f)} df,
 \end{equation*}

where $S_n(f)$ is the power spectral density period.\\

In~\cref{fig-interpolation,fig-sample-rate,fig-acceleration}, we study how this mismatch varies with interpolation order, sampling rate and LOS acceleration. We approximate the continuum limit by generating the waveform at a very fine sampling resolution ($dt = 1/262144$).  In these figures, we demonstrate these comparisons with fiducial GW170817-like BNS, GW200105-like NSBH, and GW150914-like BBH signals using base-waveform models \texttt{IMRPhenomD\_NRTidal} \cite{Dietrich:2019kaq}, \texttt{IMRPhenomNSBH} \cite{Thompson:2020nei}, and \texttt{IMRPhenomXPHM} \cite{Pratten:2020ceb}, respectively. We evaluate mismatches using the \texttt{aLIGOZeroDetHighPower} PSD with a 20 Hz lower frequency cutoff.
We use this to determine the interpolation order and sampling rate required to ensure that the resulting mismatch is smaller than other sources of uncertainty, including both statistical errors
and waveform systematic effects \citep{Cutler:2007mi, Lindblom:2008cm}.

\section{Measuring Acceleration} \label{III}

To measure LOS acceleration, we perform Bayesian Inference \cite{Thrane_2019, Rover:2006ni} using the \texttt{PyCBC Inference} toolkit \cite{Biwer:2018osg}. We analyze strain data from LIGO Hanford, LIGO Livingston, and Virgo detectors \cite{LIGOScientific:2025snk, KAGRA:2023pio} from their first (O1) to first part of fourth (O4a) observing runs as well as a small subset of data from subsequent observing runs. We generate template waveforms by directly applying the Doppler transformation, as described in section \ref{II} to different base waveform models. A list of base-waveform models for different sources is in table \ref{tab-source-base-model}.

\begin{table*}
\centering
\begin{tabular}{lccccccc}
\hline
\addlinespace
\textbf{Source Type} & \textbf{Event} & \textbf{Base-waveform model} \\
\addlinespace
\hline
\addlinespace
\addlinespace
BBH & all other events & \texttt{IMRPhenomXPHM} \cite{Pratten:2020ceb} / \texttt{IMRPhenomXO4a} \cite{Thompson:2023ase} \\
  & GW190814 & \texttt{IMRPhenomXO4a} \cite{Thompson:2023ase}, \texttt{SEOBNRv5PHM} \cite{Ramos-Buades:2023ehm}\\
  & GW250114 & \texttt{SEOBNRv5PHM} \cite{Ramos-Buades:2023ehm}\\
  & GW230814 & \texttt{SEOBNRv5PHM} \cite{Ramos-Buades:2023ehm} \\
  & GW241011 & \texttt{SEOBNRv5PHM} \cite{Ramos-Buades:2023ehm} \\
\hline
NSBH & GW200115, GW230518, GW230529 & \texttt{IMRPhenomNSBH} \cite{Thompson:2020nei} \\
     & GW200105 & \texttt{IMRPhenomNSBH} \cite{Thompson:2020nei}, \texttt{SEOBNRv5HM} \cite{Pompili:2023tna}, \texttt{SEOBNRv5EHM} \cite{Gamboa:2024hli}\\
\hline
BNS & GW170817, GW190425 & \texttt{IMRPhenomPv2\_NRTidal} \cite{Dietrich:2018uni}, \texttt{IMRPhenomD\_NRTidal} \cite{Dietrich:2019kaq} \\
\addlinespace
\hline
\end{tabular}
\caption{Base waveform models used for different types of binary systems. For a few systems, we used broader range of waveform models. For example, in the case of GW200105, we use both eccentric and non-eccentric waveform models, whereas for BNS systems we used waveform models which account for tidal deformabilities.}
\label{tab-source-base-model}
\end{table*}

We choose standard priors for various intrinsic and extrinsic parameters. The prior distributions for the common parameters which are used in all the PE runs are listed in table \ref{table-priors}. 
\begin{table*}[]
\caption{The priors used in the analysis of sources in this study}
\label{table-priors}
\begin{tabular}{|l|l|}
\hline
\textbf{Name} & \textbf{Prior Type} \\
\hline
chirp-mass, $\mathcal{M}$ & uniform \\
\hline
mass-ratio, $q$ & uniform \\
\hline
spin vectors $(\vec{S}_1, \vec{S}_2)$ & isotropic distribution for direction and uniform for amplitude \\
\hline
coalescence time, $t_c$ (s) & uniform \\
\hline
ra/dec $(\alpha/\delta)$ & isotropic sky location \\
\hline
inclination/polarization/coa\_phase, $(\iota/\psi/\phi_c)$ & isotropic binary orientation \\
\hline
luminosity distance, $D_L$ (Mpc) & uniform in volume \\
\hline
LOS acceleration, acc (cs$^{-1}$) & uniform \\
\hline
tidal deformabilities $(\Lambda_1, \Lambda_2)$ & uniform \\
\hline
\end{tabular}
\end{table*}

For NSBH and BNS events, additional parameters for tidal deformabilities are used. We choose a uniform distribution for tidal deformability. We use 20 Hz as our reference frequency to analyze all the events. Starting frequencies for BBH, NSBH, and BNS PEs are 15, 20, and 20 Hz, respectively.  
 To sample the posterior distribution, we use the \texttt{dynesty} \cite{Speagle_2020} nested sampler. We test for convergence by comparing against different numbers of live points. We analyze sufficient data to make sure that all included modes of every respective waveform models are fully encapsulated within the data.

\section{Constraining LOS acceleration of current binaries}\label{IV}

\begin{figure*}
    \begin{center}
 \includegraphics[width=\linewidth]{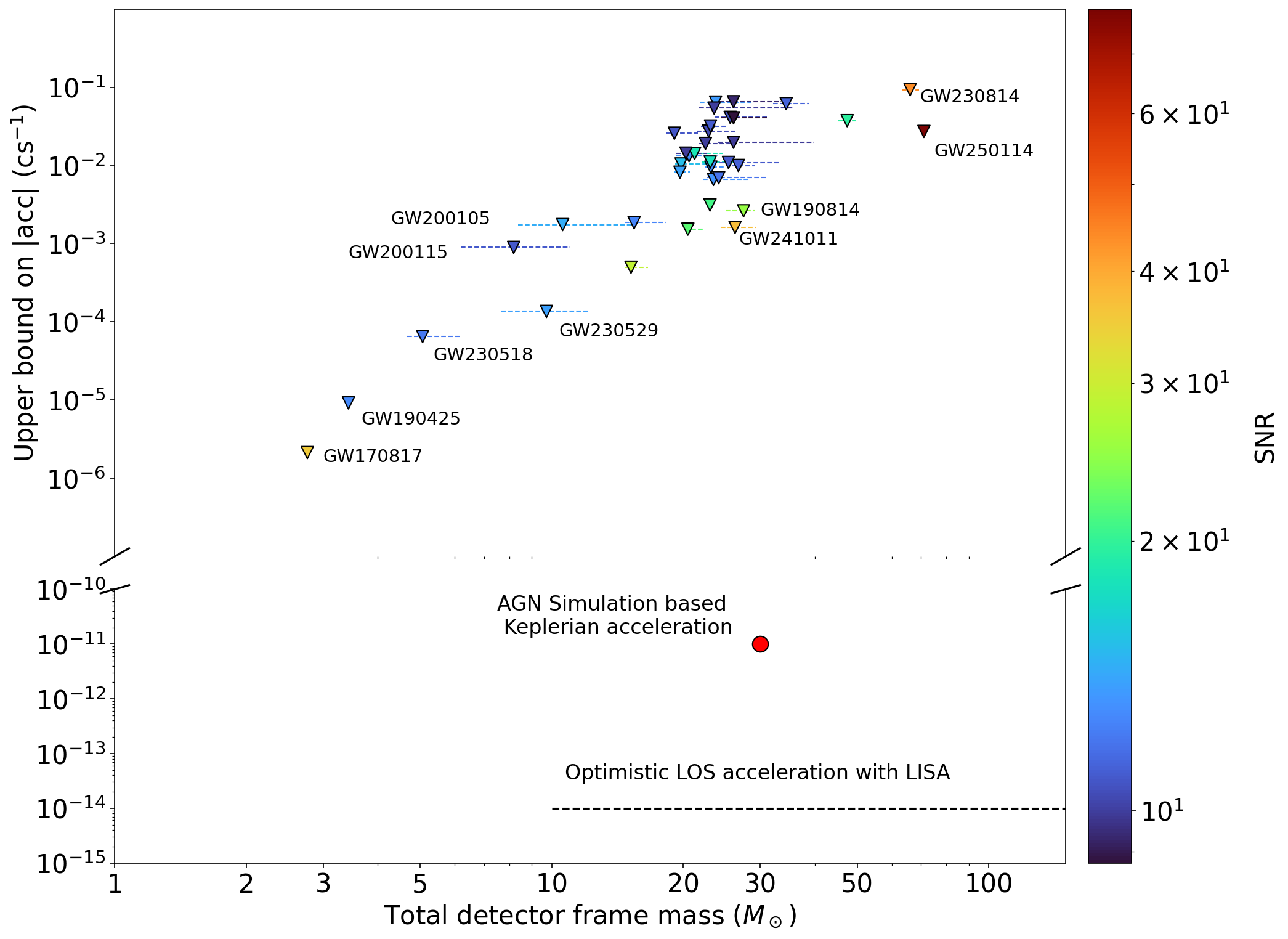}
    \caption{The upper bounds on $|acc|$ ($\textrm{c}s^{-1}$) for current GW sources. The vertical and horizontal axes show the $90\%$ credible upper bound on the LOS acceleration and the total detector frame mass, respectively, for each observation. We include only sources for which meaningful acceleration constraints can be obtained, i.e., systems that are not affected by waveform truncation or tapering systematics. This excludes the majority of BBH mergers. Horizontal dashed lines of each event show the $90\%$ credible region in total mass for each event. For comparison, we show a representative astrophysical scenario in which a binary orbits a $10^8 M_{\odot}$ SMBH at a separation of 0.045 pc, producing a Keplerian acceleration (red circle) \cite{McKernan:2024kpr}. While current ground-based detectors are not sensitive to accelerations of this magnitude, such environments may become accessible to future space-based observatories like Laser Interferometer Space Antenna (LISA). The black dashed line indicates the approximate acceleration scale at which such effects may become detectable with LISA.} 
    \label{fig-mass-vs-acc-result}
    \end{center}
\end{figure*}

We present the acceleration constraints for all the available ground-based observations through O4a and some selective observations from O4b and O4c. LOS acceleration constraints from all the observations are consistent with zero.  Fig.~\ref{fig-mass-vs-acc-result} shows how the acceleration constraints depend on the mass spectrum and signal-to-noise ratio (SNR) of each source. BNS and NSBH observations yield the tightest constraints on LOS acceleration, with progressively weaker constraints for higher-mass systems. This trend arises because lower-mass binaries spend more cycles in the sensitive band of ground-based detectors, making them more susceptible to measurable phase modifications induced by LOS acceleration. Although for most BBHs, we cannot actually obtain LOS acceleration constraints, the loudest signals like GW250114 \cite{LIGOScientific:2025wao, LIGOScientific:2025rid} make this possible. This trend can be understood from a rough scaling in which the LOS acceleration sensitivity improves approximately as $\sim 1/(T^2 \rho)$
, where $T$ and $\rho$ are the signal duration and SNR, respectively.

\subsection{BBH}  Fig.~\ref{fig-mass-vs-acc-result} includes only those BBH events for which the acceleration posterior can be constrained without being dominated by the non-relativistic truncation criterion. The excluded events have comparable or smaller durations but lower SNRs, resulting in broader acceleration posteriors that extend into regions where the assumptions underlying our waveform model are no longer valid. 

\subsubsection{GW190814}
GW190814 is a BBH merger consistent with the coalescence of a $22.2 - 24.3 M_{\odot}$ black hole and a $2.50-2.67 M_{\odot}$ compact object, indicating a significantly asymmetric mass ratio, detected with a signal-to-noise ratio (SNR) of 25 \cite{LIGOScientific:2020zkf}. The event contains measurable higher-order mode contributions and has attracted considerable interest because of its unusual mass configuration. Several studies have suggested formation channels involving dynamical environments and hierarchical triple systems \cite{LIGOScientific:2020zkf, ArcaSedda:2021zmm, Lu:2020gfh}. \citet{Yang_2025} reported evidence for nonzero LOS acceleration using a waveform model including 3.5PN acceleration corrections. However, subsequent studies \cite{Hendriks:2026kys, Pathak:2026cik, Roy:2026mco} do not confirm this finding and instead show consistency with zero acceleration within uncertainties.
\begin{figure} []
\includegraphics[width=8cm,height=6cm,angle=0]{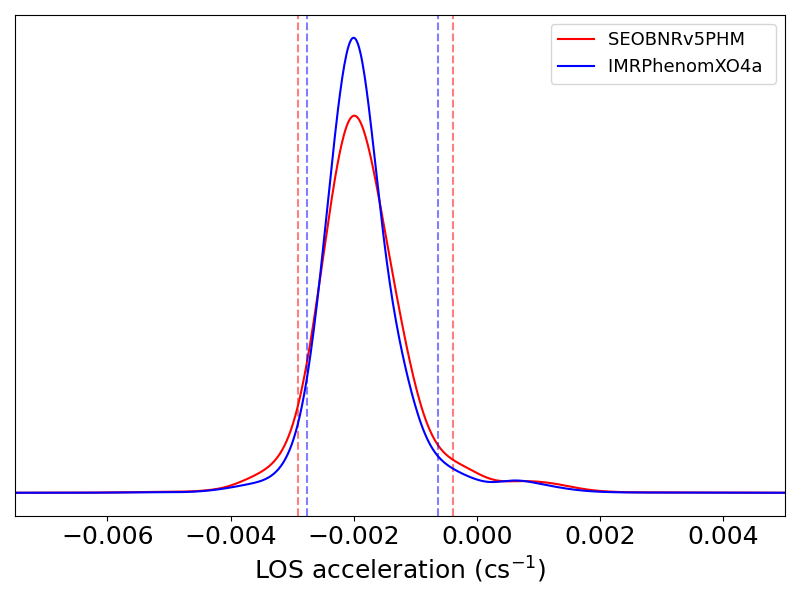}
\caption{LOS acceleration for GW190814 using  \texttt{SEOBNRv5PHM}(red) and \texttt{IMRPhenomXO4a}(blue). Dashed vertical lines show the $90 \%$ credible region. LOS acceleration for GW190814 is consistent with zero, which is different from the previously claimed non-zero LOS acceleration \cite{Yang_2025}. Our posterior is in good agreement with \citet{Roy:2026mco}. \citet{Hendriks:2026kys} also found no compelling evidence for non-zero LOS acceleration, although differences in the inferred posteriors likely reflect differences in waveform modeling and the treatment of higher-order modes.}
\label{fig-GW190814}
\end{figure}
Figure~\ref{fig-GW190814} shows the posteriors on the LOS acceleration obtained using the \texttt{SEOBNRv5PHM} \cite{Ramos-Buades:2023ehm}  and \texttt{IMRPhenomXO4a} \cite{Thompson:2023ase} base waveform models. We find no evidence for non-zero LOS acceleration. While the posteriors are slightly offset from zero, the Bayes factor computed using the Savage-Dickey density ratio (SDDR) \cite{Dickey1971TheWL} for \texttt{SEOBNRv5PHM} \cite{Ramos-Buades:2023ehm}  and \texttt{IMRPhenomXO4a}\cite{Thompson:2023ase} are 1.9 and 2.8, respectively, indicating no support for acceleration.

\subsubsection{GW241011}
\begin{figure} []
\includegraphics[width=8cm,height=6cm,angle=0]{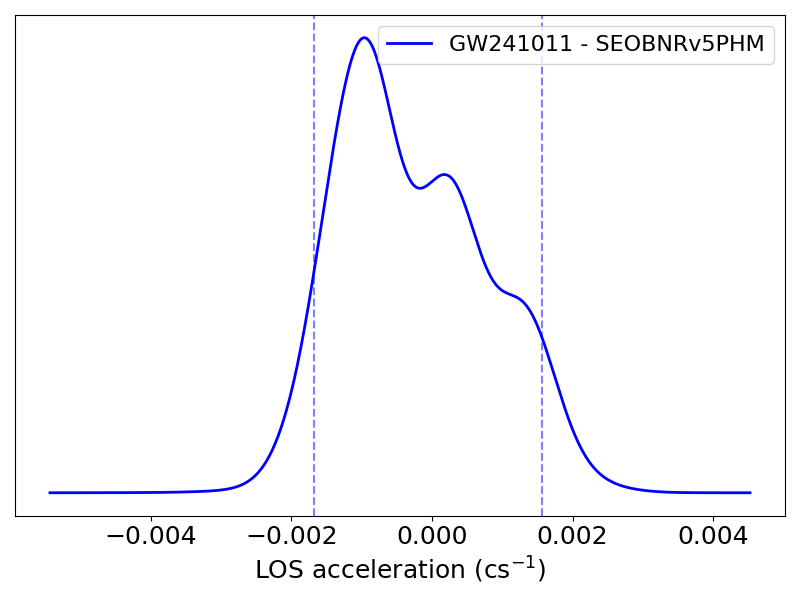}
    \caption{The LOS acceleration for GW241011. The measurement is consistent with zero LOS acceleration.}
    \label{fig-GW241011}
\end{figure}

GW241011 is a BBH merger consistent with component masses of $19.6_{-2.5}^{+3.6} M_{\odot}$ and $5.9_{-0.8}^{+0.8} M_{\odot}$ black holes, indicating an asymmetric mass ratio. It was detected with an SNR of 35.8 and has a high primary spin magnitude of $0.78_{-0.9}^{+0.9}$ \cite{LIGOScientific:2025brd}. We analyze this event using \texttt{SEOBNRv5PHM} \cite{Ramos-Buades:2023ehm}. The acceleration probability distribution is shown in \ref{fig-GW241011}. The measurement is consistent with zero, with a median LOS acceleration of $-4.4_{-12.3}^{+20.0}\times10^{-4} \,\textrm{c}s^{-1}$.

\subsubsection{GW250114} 
\begin{figure} []
\includegraphics[width=8cm,height=6cm,angle=0]{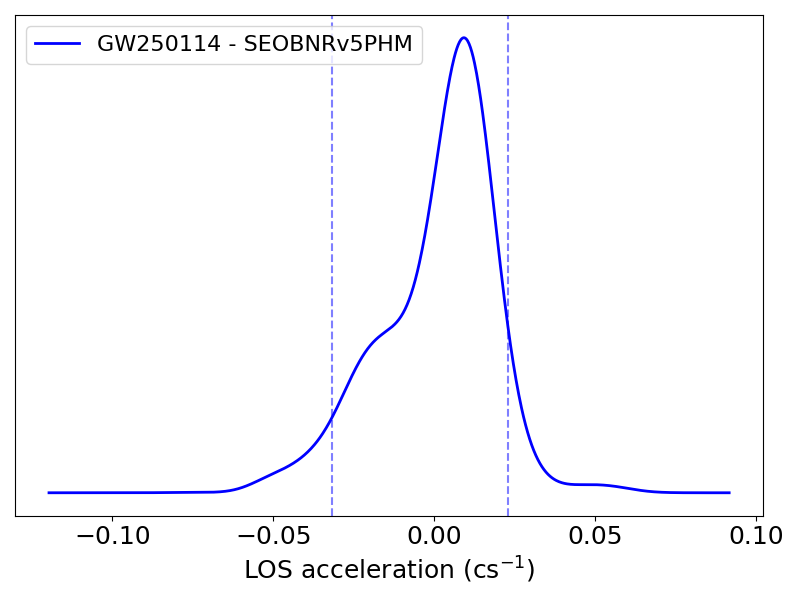}
    \caption{The LOS acceleration for GW250114. This is one of the few BBHs where LOS acceleration constraints can be obtained without begin dominated by the non-relativistic truncation.}
    \label{fig-GW250114}
\end{figure}

GW250114 is the loudest GW event with an SNR of approximately 80 detected to date \cite{LIGOScientific:2025rid, LIGOScientific:2025wao, Akyuz:2025seg}. Its component masses are similar to those of GW150914 \cite{LIGOScientific:2016vbw}. We analyse this event using  \texttt{SEOBNRv5PHM} \cite{Ramos-Buades:2023ehm} waveform model. The acceleration probability distribution is shown in Fig.~\ref{fig-GW250114}. The high SNR yields tighter acceleration constraints than those obtained for BBH systems of comparable mass. Nevertheless, the measurement is consistent with zero, with an acceleration found to be $5_{-37}^{+18}\times10^{-3} \,\textrm{c}s^{-1}$ at $90\%$ credible level.  

\subsection{NSBH} 
Currently, there have been four observed NSBH events. Their lower component masses produce substantially tighter acceleration constraints than those obtained for BBH systems. For GW200115 \cite{Mandel:2021ewy}, GW230518 \cite{LIGOScientific:2025pvj}, and GW230529 \cite{Sanger:2024axs}, we use \texttt{IMRPhenomNSBH} \cite{Thompson:2020nei} as the base waveform model. 

\begin{figure} 
    \begin{center}
\includegraphics[width=8cm,height=6cm,angle=0]{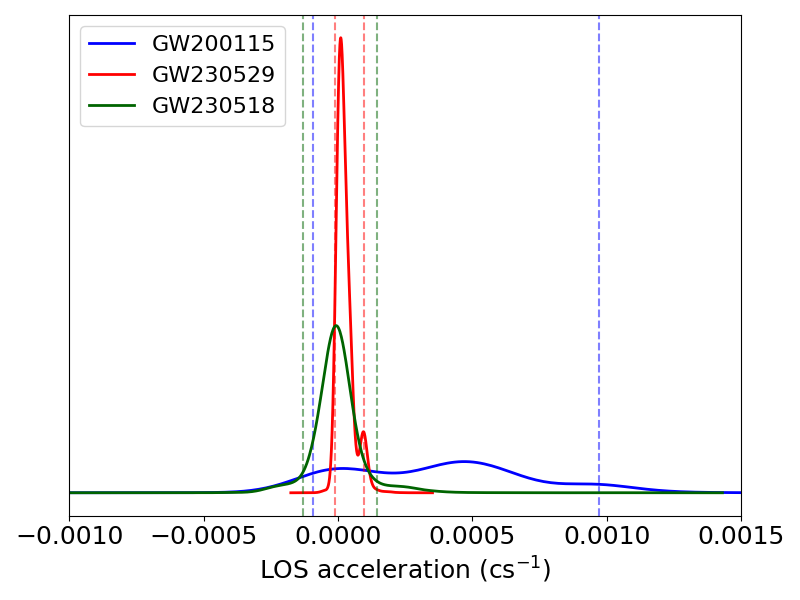}
    \caption{LOS acceleration distribution for all the NSBH events except GW200105. These are all consistent with zero LOS acceleration.}
    \label{fig-NSBH-acc}
    \end{center}
\end{figure} 
Our measurements are consistent with zero, with $90 \%$ credible intervals of $+4.0_{-5.0}^{+5.7}\times 10^{-4} \,\textrm{c}s^{-1}$ for GW200115, $+1.7_{-2.7}^{+7.8}\times 10^{-5
} \,\textrm{c}s^{-1}$ for GW230529 and $-6.8_{-120}^{+150}\times 10^{-6} \,\textrm{c}s^{-1}$ for GW230518. The particularly tight constraint obtained for GW230529 is primarily driven by its longer duration, while the modest improvement of GW230518 relative to GW200115 is attributable to its higher SNR.

\subsubsection{GW200105} 
Among the NSBH events, GW200105 \cite{LIGOScientific:2021qlt} is claimed to be eccentric \cite{Jan:2025fps, Fei:2024ruj, Planas:2025plq, Morras:2025xfu, Kacanja:2025kpr}. This GW event is consistent with the coalescence of a $8.9_{-1.5}^{+1.2} M_{\odot}$ black hole and a $1.9_{-0.2}^{+0.3} M_{\odot}$ neutron star, detected in a single detector with a SNR of 13.2 \cite{LIGOScientific:2021qlt}. We analyse this event with both non-eccentric (\texttt{IMRPhenomNSBH} \cite{Thompson:2020nei}, \texttt{SEOBNRv5HM} \cite{Pompili:2023tna}) and eccentric (\texttt{SEOBNRv5EHM} \cite{Gamboa:2024hli}) waveform models. When this event is analysed with non-eccentric waveform models, the posterior excludes zero acceleration at the $90\%$ credible level. The corresponding credible interval with \texttt{SEOBNRv5HM} \cite{Pompili:2023tna} is $-1.36_{-0.31}^{+1.04}\times 10^{-3} \,\textrm{c}s^{-1}$. After considering the eccentric waveform model, the resulting posterior is consistent with zero acceleration, with a $90 \%$ credible interval of $-6.5_{-8.5}^{+22}\times 10^{-4} \,\textrm{c}s^{-1}$. The resulting posteriors are shown in Fig~\ref{fig-acceleration}.

    \begin{figure} []
    \begin{center}
\includegraphics[width=8cm,height=6cm,angle=0]{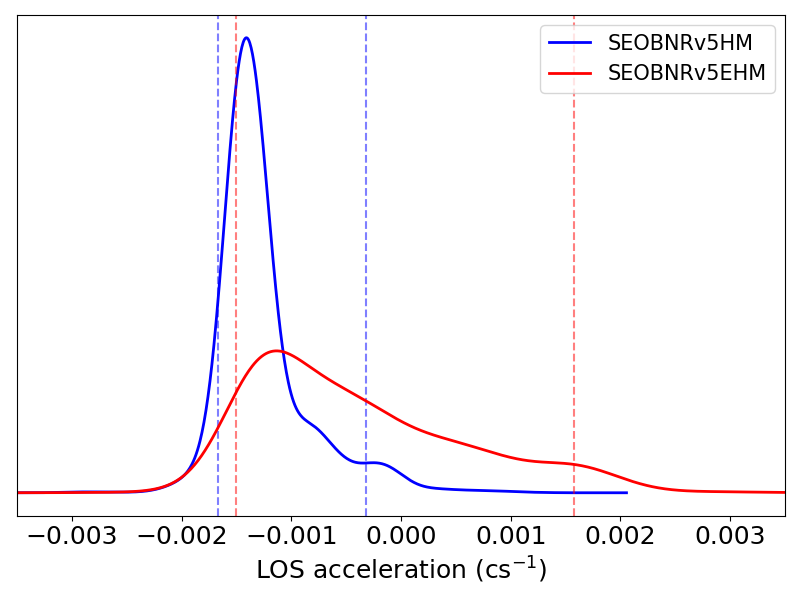}
    \caption{LOS acceleration posterior for GW$200105$ for different base waveform mdoels that neglects eccentricity (blue) and includes eccentricity (red). When eccentricity is not accounted for, the distribution appears to prefer non-zero acceleration. However, when eccentricicity is included, LOS acceleration posterior becomes consistent with zero.}
    \label{fig-GW200105-acc}
    \end{center}
\end{figure}

\begin{figure} []
    \begin{center}
\includegraphics[width=0.48\textwidth,height=8cm]{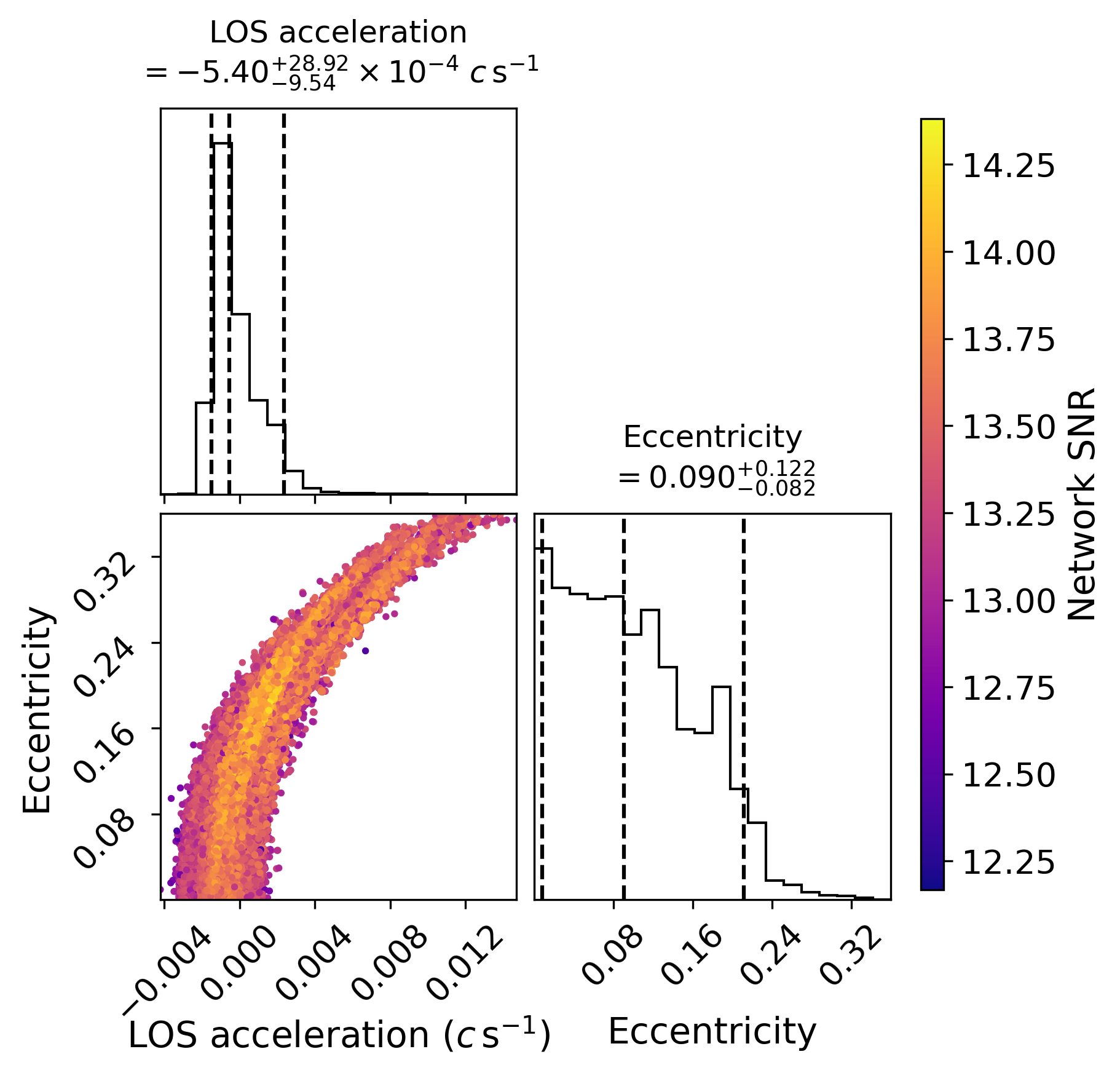}
    \caption{Marginalized posterior for GW200105 as a function of LOS acceleration and eccentricity  using the \texttt{SEOBNRv5EHM} waveform model. A correlation between LOS acceleration and eccentricity is clearly evident and it is consistent with how they qualitatively affect the waveform model.}
    \label{fig-GW200105-ecc-acc}
    \end{center}
\end{figure}

\begin{figure} []
    \begin{center}
\includegraphics[width=8cm,height=6cm,angle=0]{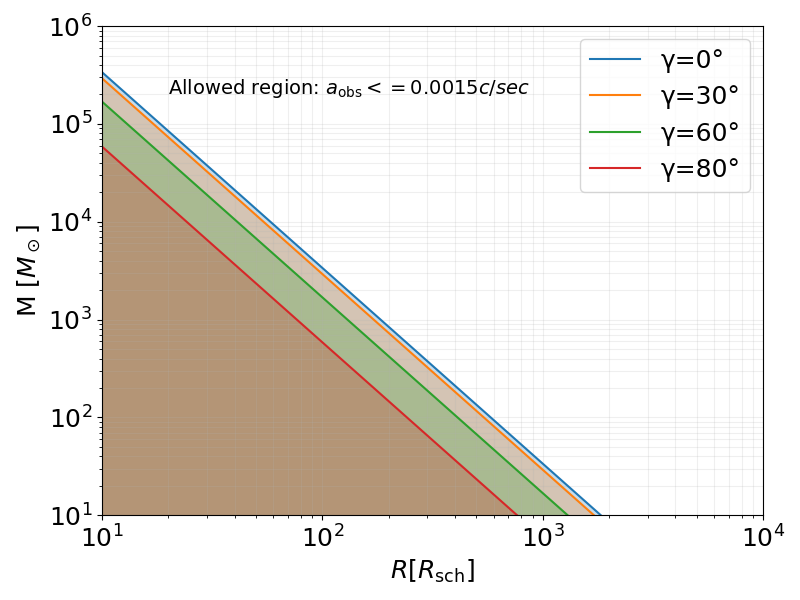}
    \caption{Constraints on the properties of a third body in the parameter space of it's mass $M\,(M_{\odot})$ and distance from binary $R\,(R_{sch})$. These constraints are derived from the observed bound on LOS acceleration. Because LOS acceleration is a function of the viewing angle, the resulting constraints also depend on the assumed viewing angle. We show the constraints for a set of viewing angles. For $90^{\circ}$ viewing angle, no constraints can be obtained.}
    \label{fig-M-R-space}
    \end{center}
\end{figure} 
To quantify the statistical preference for acceleration, we compute the Bayes factor using the SDDR \cite{Dickey1971TheWL}. This compares the zero-acceleration hypothesis against the acceleration model via the ratio of the posterior to prior density evaluated at zero acceleration. Using the \texttt{SEOBNRv5HM} \cite{Pompili:2023tna} waveform model, we obtain a Bayes factor of 9.16, indicating moderate evidence in favor of the zero-acceleration hypothesis. Using the \texttt{SEOBNRv5EHM} \cite{Gamboa:2024hli} waveform model for GW200105, we obtain a Bayes factor of 25.32, indicating evidence against acceleration and in favor of eccentricity.

The joint posterior distribution shown in Fig.~\ref{fig-GW200105-ecc-acc} reveals a clear correlation between eccentricity and LOS acceleration. In particular, regions of higher eccentricity are associated with larger positive acceleration, suggesting that the signal can be interpreted as either eccentric or accelerating when one of these effects is neglected. This behavior can be understood from the impact of both parameters on the waveform evolution. Increased eccentricity tends to accelerate the orbital evolution. However, a positive LOS acceleration redshifts the signal and slows the observed frequency evolution. Consequently, a larger positive acceleration can compensate for the more rapid evolution induced by higher eccentricity, leading to the observed correlation. \citet{Roy:2026mco} also reported a strong degeneracy between eccentricity and LOS acceleration in GW200105 and found no evidence for non-zero LOS acceleration using an eccentric waveform model.

We obtain an acceleration upper bound of approximately $0.0015\,\textrm{c}s^{-1}$. This limit can be translated into constraints on the mass and separation of a potential third body assuming Keplerian acceleration. Fig.~\ref{fig-M-R-space} shows the corresponding allowed regions in the $M$-$R$ parameter space for different viewing angles $\gamma$ between the binary and third body separation vector and the LOS vector. The contours corresponding to different viewing angles $\gamma$ exclude the regions below them, whereas the regions above remain consistent with the acceleration constraint. As $\gamma$ increases, the projected LOS acceleration decreases, and a smaller region is ruled out. In the limit $\gamma=90^{\circ}$, the projected acceleration vanishes and the entire parameter space remains consistent with the measurement.

\subsection{BNS} 
We perform the BNS analysis using \texttt{IMRPhenomPv2\_NRTidal} \cite{Dietrich:2018uni} as the baseline waveform model. Both GW170817 \cite{LIGOScientific:2017vwq} and GW190425 \cite{LIGOScientific:2020aai} are consistent with zero LOS acceleration. BNS systems provide the tightest acceleration constraints in the current catalog owing to their longer durations. For GW190425 and GW170817, we obtain a $90\%$ credible interval of $-3.3_{-7.5}^{+7.3}\times 10^{-6} \,\textrm{c}s^{-1}$ and $+1.0_{-1.4}^{+1.4}\times 10^{-6} \,\textrm{c}s^{-1}$, respectively. Despite differences in the Doppler transformation methodology and modeling approach, our results are broadly consistent with those of \citet{Vijaykumar:2023tjg}, who employed a waveform model incorporating 3.5PN acceleration corrections.
\begin{figure} []
    \begin{center}
\includegraphics[width=8cm,height=6cm,angle=0]{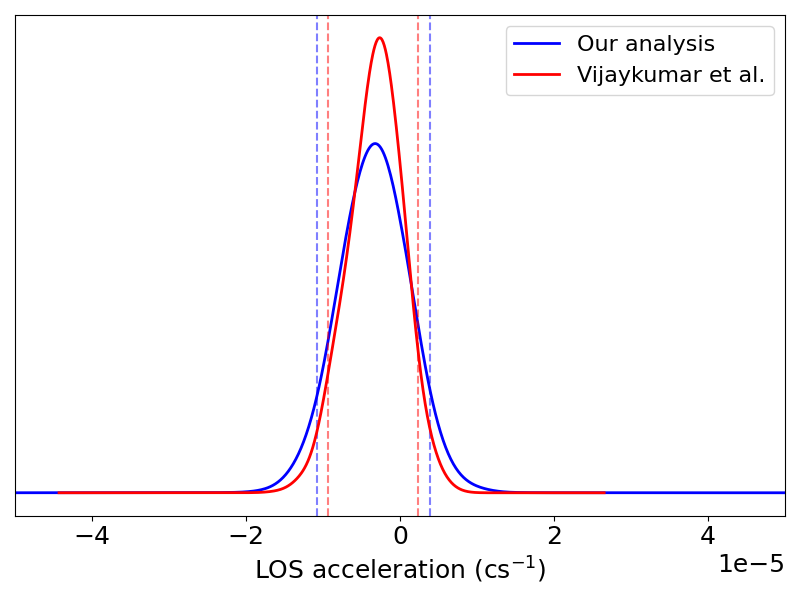}
    \caption{LOS acceleration posterior for GW$190425$. Our result (blue) is consistent with zero LOS acceleration. Our result is broadly consistent with the previous analysis from \citet{Vijaykumar:2023tjg} (red).}
    \label{fig-GW190425}
    \end{center}
\end{figure} 

\begin{figure} []
    \begin{center}
\includegraphics[width=8cm,height=6cm,angle=0]{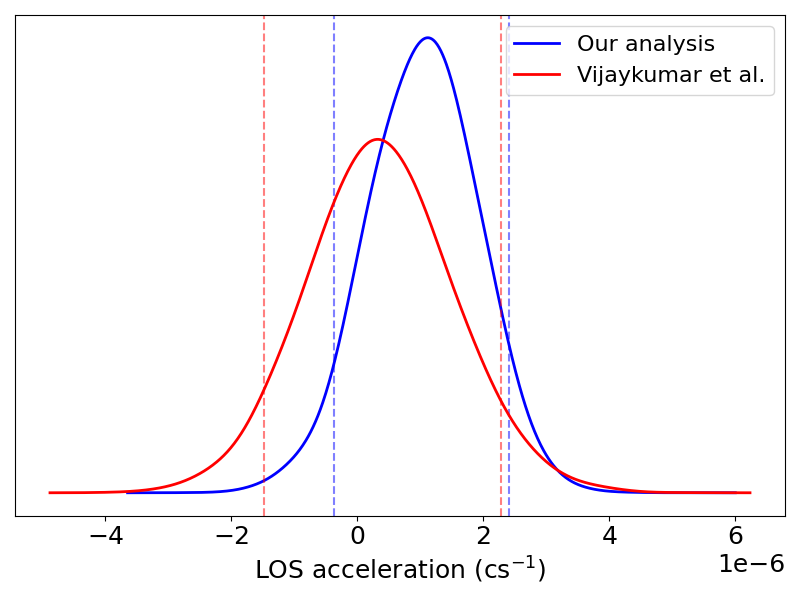}
    \caption{LOS acceleration posterior for GW$170817$. Our result (blue) is consistent with zero LOS acceleration. Our result is broadly consistent with the previous analysis from \citet{Vijaykumar:2023tjg} (red).}
    \label{fig-GW170817}
    \end{center}
\end{figure} 

\balance

\section{Summary and Discussion}\label{V}
We measured the LOS acceleration for all observed sources through O4a and select sources from later observing runs. We introduced a new method to model the time-varying Doppler shift by directly applying the tranformaion in the time-domain to an arbitrary underlying waveform model, including those with higher order modes, precession, eccentricity, etc. Similar to prior studies, we have assumed a constant LOS acceleration. Our results are consistent with zero LOS acceleration for all observations. LOS acceleration for BNS signals, which have much longer duration, can be constrained down to $10^{-6} \,\textrm{c}s^{-1}$, whereas even the best BBH events, such as GW250114, can only be constrained down to $10^{-2} \,\textrm{c}\textrm{s}^{-1}$. This difference is consistent with the differences in SNR and signal duration. Our results have demonstrated the correlation between eccentricity and LOS acceleration as evident by GW200105. Previous studies \cite{Jan:2025fps, Fei:2024ruj, Planas:2025plq, Morras:2025xfu, Kacanja:2025kpr} have interpreted this event as being eccentric. We find that if we account for eccentricity, this event is consistent with zero LOS acceleration.

Our approach of directly applying the time varying Doppler transformation in the time domain enables the use of more general LOS velocity profiles than a constant acceleration approximation. However, it still assumes uniform acceleration throughout the entire binary, neglecting any internal dynamical or tidal effects. For sufficiently tight binaries or longer duration signals, higher order variations in the LOS velocity become important, and a constant acceleration description is no longer adequate ~\cite{Ransom:2002cj}.

There is growing evidence that some of the most massive GW events, such as GW190521 \cite{LIGOScientific:2020iuh} and GW231123 \cite{LIGOScientific:2025rsn}, originate from dense dynamical environments, including AGN disks and hierarchical systems \cite{Delfavero:2025lup, Angeloni:2026nmy, Holgado:2020imj, Romero-Shaw:2020thy, DallAmico:2021hze, Graham:2020gwr}. Compact binaries formed in such environments are expected to experience non-negligible LOS accelerations due to the gravitational potential of nearby massive objects. Figure~\ref{fig-mass-vs-acc-result} compares the LOS acceleration expected for a representative AGN scenario \cite{McKernan:2024kpr}, in which a $30 M_\odot$ binary orbits a $10^8 M_\odot$ SMBH, with the constraints obtained from current observations. The expected astrophysical accelerations lie well below the sensitivity of present ground-based detectors. 

Next generation ground based observatories \cite{Reitze:2019iox, ET:2025xjr} are expected to improve these constraints substantially. Their increased strain sensitivity will provide higher SNRs, while their improved low-frequency sensitivity will allow binaries to be observed for much longer durations. Since the phase shift induced by LOS acceleration accumulates throughout the inspiral, longer observations will significantly enhance sensitivity to this effect. Consequently, next-generation detectors are expected to improve acceleration constraints by several orders of magnitude, potentially reaching the regime relevant for some astrophysical formation scenarios \cite{Tamanini:2019usx, Vijaykumar:2023tjg, McKernan:2024kpr}. However, accurate modeling of environmental effects will become increasingly important, as neglecting LOS acceleration can introduce systematic biases in the estimation of intrinsic binary parameters, such as the chirp mass and symmetric mass ratio, particularly for low-mass and asymmetric mass-ratio binaries \cite{Gera:2025ugl}.

Space-based detectors such as LISA \cite{Buchman:2013msh} offer an even more promising avenue for measuring LOS acceleration. The long-duration observations of stellar-mass binaries will allow tiny Doppler-induced phase drifts to accumulate over several years, bringing the expected sensitivity close to the LOS accelerations predicted for AGN-assisted mergers \cite{Tamanini:2019usx}. Moreover, although LOS acceleration and orbital eccentricity can exhibit partial degeneracies in current ground-based observations, they affect the waveform in fundamentally different ways. LOS acceleration induces a time varying Doppler modulation of the signal, whereas eccentricity modifies the intrinsic orbital dynamics through higher harmonics and apsidal precession. The rich low-frequency waveform observed by LISA is therefore expected to distinguish these effects more effectively, reducing parameter degeneracies. Multiband observations \cite{Sesana:2016ljz, Amaro-Seoane:2009vjl, Wu:2025zhc} combining LISA with next-generation ground-based detectors could provide particularly powerful constraints on the environments of compact binaries  \cite{Toubiana:2020drf, Tamanini:2019usx, Sberna:2022qbn}.

Code and selected posterior samples are available at \url{https://github.com/labani-01/LOS-acceleration_LIGO-pe}; additional data products are available upon request.

\section{Acknowledgements}
LR and AHN acknowledge the support from Syracuse University for providing computational resources through the sugwg High Throughput Computing (HTC) cluster supported by NSF award ACI-1341006. AHN acknowledges support from NSF grant PHY-2309240. LR is also grateful to Aditya Vijaykumar for sharing their results, which were used for comparison in this study. LR would also like to thank Collin Capano, Keisi Kacanja, Kanchan Soni, Alex Correia, Vikas Jadhav Y, and Aleyna Akyüz for helpful discussions and feedback throughout this work, and for valuable input on various aspects of the analysis. LR further thanks Ritapriya Pradhan for many insightful discussions throughout the course of this work. 
 
This research has used data or software obtained from the Gravitational Wave Open Science Center (gwosc.org), a service of the LIGO Scientific Collaboration, the Virgo Collaboration, and KAGRA. This material is based upon work supported by NSF's LIGO Laboratory which is a major facility fully funded by the National Science Foundation, as well as the Science and Technology Facilities Council (STFC) of the United Kingdom, the Max-Planck-Society (MPS), and the State of Niedersachsen/Germany for support of the construction of Advanced LIGO and construction and operation of the GEO600 detector. Additional support for Advanced LIGO was provided by the Australian Research Council. Virgo is funded, through the European Gravitational Observatory (EGO), by the French Centre National de Recherche Scientifique (CNRS), the Italian Istituto Nazionale di Fisica Nucleare (INFN) and the Dutch Nikhef, with contributions by institutions from Belgium, Germany, Greece, Hungary, Ireland, Japan, Monaco, Poland, Portugal, Spain. KAGRA is supported by Ministry of Education, Culture, Sports, Science and Technology (MEXT), Japan Society for the Promotion of Science (JSPS) in Japan; National Research Foundation (NRF) and Ministry of Science and ICT (MSIT) in Korea; Academia Sinica (AS) and National Science and Technology Council (NSTC) in Taiwan.

\bibliographystyle{apsrev4-1}
\bibliography{ref}

\end{document}